\documentclass[twocolumn,aps,prl,showpacs,superscriptaddress]{revtex4}
\usepackage{graphicx}

\begin{document}

\author{Bret Heinrich}
\affiliation{Department of Physics, Simon Fraser University, Burnaby, BC, V5A 1S6, Canada}
\author{Yaroslav Tserkovnyak}
\affiliation{Lyman Laboratory of Physics, Harvard University, Cambridge, Massachusetts 02138}
\author{Georg Woltersdorf}
\affiliation{Department of Physics, Simon Fraser University, Burnaby, BC, V5A 1S6, Canada}
\author{Arne Brataas}
\affiliation{Department of Physics, Norwegian University of Science and Technology, N-7491 Trondheim, Norway}
\author{Radovan Urban}
\affiliation{Department of Physics, Simon Fraser University, Burnaby, BC, V5A 1S6, Canada}
\author{Gerrit E. W. Bauer}
\affiliation{Department of NanoScience, Delft University of Technology, 2628 CJ Delft, The Netherlands}

\title{Dynamic Exchange Coupling in Magnetic Bilayers}

\begin{abstract}
A long-ranged dynamic interaction between ferromagnetic
films separated by normal-metal spacers is reported, which is
communicated by nonequilibrium spin currents. It is measured by
ferromagnetic resonance (FMR) and explained by an adiabatic
spin-pump theory. In FMR the spin-pump mechanism of
spatially separated magnetic moments leads to an appreciable
increase in the FMR line width when the resonance fields are well
apart, and results in a dramatic line-width narrowing when the FMR
fields approach each other.
\end{abstract}

\date{\today}
\maketitle


The giant magnetoresistance \cite{Baibich:prl88} accompanying realignment of magnetic configurations in metallic
 multilayers by an external magnetic field is routinely employed in magnetic read heads and is essential
 for high-density nonvolatile magnetic random-access memories. These typically consist of
  ferromagnetic/normal/ferromagnetic (\textit{F{\rm /}N{\rm /}F}) metal hybrid structures, i.e.,  magnetic bilayers which are an essential building block of the so called spin valves.
   The static  Ruderman-Kittel-Kasuya-Yosida (RKKY) interlayer exchange between ferromagnets in magnetic multilayers \cite{Grunberg:prl86} is suppressed in these devices by a sufficiently thick nonmagnetic spacer \textit{N} or a tunnel barrier. The interest of the community shifts increasingly from the
static to the dynamic properties of the magnetization
\cite{Back:sc99}. This is partly motivated by curiosity, partly by the
fact that the magnetization switching characteristics in memory
devices is a real technological issue. A good grasp of the fundamental
physics of the magnetization dynamics becomes of essential importance
to sustain the exponential growth of device performance factors.

In this Letter we study the largely unexplored
dynamics of magnetic bilayers in a regime when there is no discernible \textit{static} interaction between the magnetization vectors.
Surprisingly, the magnetizations still turn
out to be coupled, which we explain by emission and absorption of
nonequilibrium spin currents.
Under special conditions the two magnetizations are resonantly
coupled by spin currents and carry out a synchronous motion, quite analogous to two connected pendulums. This
dynamic interaction is an entirely new concept and physically very
different from the static RKKY coupling.
E.g., the former does not oscillate as a function of thickness
and its range is exponentially limited by the spin-flip relaxation length of spacer layers and algebraically by the elastic mean
 free path. This coupling can have profound effects on magnetic relaxation and switching behavior in hybrid structures and devices.

\begin{figure}
\includegraphics[width=8.6cm,clip=]{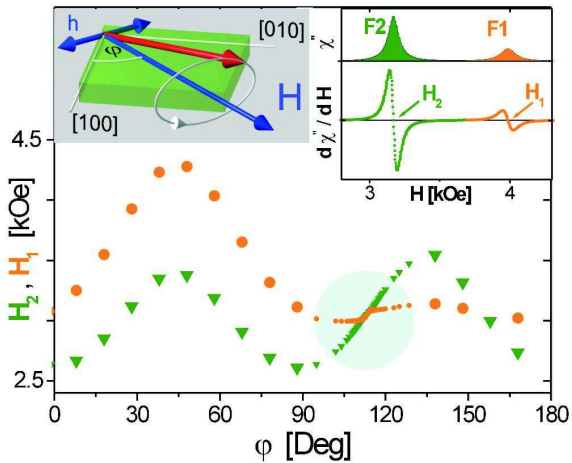}
\caption{\label{fig1} Dependence of the FMR resonance fields $H_1$
(circles) and $H_2$ (triangles) for the thin Fe film
\textit{F1}, and the thick Fe film \textit{F2}, respectively,
on the angle $\varphi$ of the external dc magnetic field with respect
to the Fe [100] crystallographic axis.
The sketch of the in-plane measurement in the left inset shows how the rf magnetic field (double-pointed arrow) drives the magnetization (on a scale grossly exaggerated for easy viewing). In the right inset we plot the measured
absorption peaks for layers \textit{F1} and
\textit{F2} at $\varphi=60$~Deg.}
\end{figure}

The unit vector $\mathbf{m}=\mathbf{M}/M$ of the magnetization
$\mathbf{M}(t)$ of a ferromagnet changes its direction in the
presence of a noncollinear magnetic field.
The motion of $\mathbf{m}$ in a single domain is described by the
Landau-Lifshitz-Gilbert (LLG) equation
\begin{equation}
\frac{d\mathbf{m}}{dt}=-\gamma\mathbf{m}\times\mathbf{H}_{\text{eff}}+\alpha\mathbf{m}\times\frac{d\mathbf{m}}{dt}\, ,
\label{llg}
\end{equation}
with $\gamma$ being the absolute value of the gyromagnetic ratio.
The first term on the right-hand side represents the torque
induced by the effective magnetic field
$\mathbf{H}_{\text{eff}}=-\partial F/\partial\mathbf{M}$, where
the free-energy functional $F[\mathbf{M}]$ consists of the Zeeman
energy, magnetic anisotropies, and exchange interactions
\cite{Heinrich:ap93}. The second term in Eq.~(\ref{llg}) is the
Gilbert damping torque which governs the relaxation towards
equilibrium. The intrinsic damping in bulk metallic ferromagnets,
$\alpha^{(0)}$, typically 0.002-0.025, appears to be governed by
spin-orbit interactions \cite{Kunes:prb02} in the 3\textit{d}
transition metals. The magnetization vector can be forced into a
resonant precession motion by microwave stimulation. This ferromagnetic
resonance (FMR) is measured via the absorption of microwave
power using a small rf field at a frequency $\omega$ polarized
perpendicular to the static magnetic moment as a function of the
applied dc magnetic field, see the right inset in Fig.~\ref{fig1}.
The absorption is given by the imaginary part of the
susceptibility $\chi^{\prime\prime}$ of the rf magnetization component
along the rf driving field. This FMR signal has a Lorentzian line
shape with a width $\Delta H=(2/\sqrt{3})\alpha\omega/\gamma$ when
defined by the inflection points (i.e., the extrema of
$d\chi^{\prime\prime}/dH$), see the left inset in Fig.~\ref{fig1}.

When two or more ferromagnets are in electrical contact via nonmagnetic metal layers, interesting new effects occur. Transport of spins accompanying an applied electric current driven through a magnetic
multilayer causes a torque on the magnetizations \cite{Sloncz:mmm96},
which at sufficiently high current densities leads to
spontaneous magnetization-precession and switching phenomena \cite{Myers:sc99}.
Even in the
absence of an applied charge current, spins are injected into the normal metal by a ferromagnet with moving magnetization. This causes additional magnetic damping, provided that the spin-flip relaxation rate of normal metal is high \cite{Tserkovnyak:prl021}. The
present Letter focuses on the discovery of novel dynamic effects in
\textit{F1{\rm /}N{\rm /}F2} structures in the limit when the
spin-flip scattering in \textit{N} is weak.
Let us first sketch the basic physics.
A precessing magnetization $\mathbf{m}_i$
\textquotedblleft pumps\textquotedblright\ a spin current
$\mathbf{I}_{si}^{\text{pump}}\perp\mathbf{m}_i$ into the normal metal
\cite{Tserkovnyak:prl021}. 
We focus on weakly excited magnetic bilayers close to the parallel alignment, so that
$\mathbf{I}_{si}^{\text{pump}}\perp\mathbf{m}_j$ for arbitrary
$i,j=1,2$. The spin momentum perpendicular to the
magnetization direction cannot penetrate a ferromagnetic film
beyond the (transverse) spin-coherence length,
$\lambda_{\text{sc}}=\pi/|k_{\text{F}}^\uparrow-k_{\text{F}}^\downarrow|$, which is
determined by the spin-dependent Fermi wave vectors
$k_{\text{F}}^{\uparrow,\downarrow}$ and is smaller than a nanometer
for 3\textit{d} metals \cite{Stiles:prb02}. A
transverse spin current ejected by one ferromagnet can therefore be absorbed at the interface to the neighboring ferromagnet,
 thereby exerting a torque
$\text{\boldmath$\tau$}$. Each magnet thus acts as a spin sink
which can dissipate the transverse spin current ejected by the other
layer.

The theoretical basis of this picture is the adiabatic spin-pumping
mechanism \cite{Tserkovnyak:prl021} and magnetoelectronic
circuit theory \cite{Brataas:prl00}. \textit{N} is assumed thick
enough to suppress any RKKY \cite{Grunberg:prl86}, pin-hole
\cite{Bobo:prb99}, and magnetostatic (N\'{e}el-type)
\cite{Rucker:jap95} interactions. We consider ultrathin
films with a constant magnetization vector across the film
thickness \cite{Heinrich:ap93}, which are nonetheless
thicker than $\lambda_{\text{sc}}$ and, therefore,
completely absorb transverse spin currents. In the experiments
described below, \textit{N} is thinner than the electron mean free
path, so that the electron motion inside the spacer is ballistic.
Precessing $\mathbf{m}_i$ pumps spin angular momentum
at the rate \cite{Tserkovnyak:prl021}
\begin{equation}
\mathbf{I}_{si}^{\text{pump}}=\frac{\hbar}{4\pi}g^{\uparrow\downarrow}\mathbf{m}_i\times\frac{d\mathbf{m}_i}{dt}\,,
\label{pump}
\end{equation}
where $g^{\uparrow\downarrow}$ is the dimensionless
\textquotedblleft mixing\textquotedblright\ conductance
\cite{Brataas:prl00} of the \textit{F{\rm /}N} interfaces, which
can be obtained via \textit{ab initio} calculations of the
scattering matrix \cite{Xia:prb02} or measured via the angular
magnetoresistance of spin valves \cite{Bauer:prep} as well as FMR
line widths of \textit{F{\rm /}N} and \textit{F{\rm /}N{\rm /}F}
magnetic structures \cite{Tserkovnyak:prl021,Mizukami:mmm01,Urban:prl01}.
Note that $g^{\uparrow\downarrow}$ must be renormalized for the
intermetallic interfaces considered here \cite{Bauer:prep}. We
assume identical \textit{Fi{\rm /}N} interfaces with real-valued
$g^{\uparrow\downarrow}$, as suggested by calculations for various
\textit{F{\rm /}N} combinations \cite{Xia:prb02}.
When the spacer is not ballistic, its diffuse resistance can simply be
absorbed into the value of $g^{\uparrow\downarrow}$, which should then be
interpreted as the mixing conductance of an \textit{F{\rm /}N} interface
\textit{in series} with the half of the spacer. When, furthermore,
the spacer is thicker than the spin-diffusion length, the spin-pumping
exchange between the magnetic layers becomes exponentially suppressed
with the spacer thickness \cite{Tserkovnyak:prl021}.

\begin{figure}
\includegraphics[width=8.6cm,clip=]{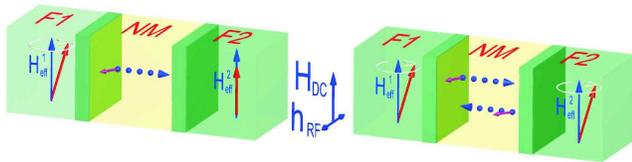}
\caption{\label{fig2} A cartoon of the dynamic coupling phenomenon.
In the left drawing, layer \textit{F1} is at a
resonance and its precessing magnetic moment pumps spin current into
the spacer, while \textit{F2} is detuned from its FMR. In the right
drawing, both films resonate at the same external field,
inducing spin currents in opposite directions. The short
arrows in \textit{N} indicate the instantaneous direction of the spin angular
momentum $\propto\mathbf{m}_i\times
d\mathbf{m}_i/dt$ carried away by the spin currents. Darker areas in
\textit{Fi} around the interfaces represent the narrow regions in
which the transverse spin momentum is absorbed.}
\end{figure}

Alloy disorder at the interfaces scrambles the distribution function.
Disregarding spin-flip scattering in the normal metal, an
incoming spin current on one side leaves the
normal-metal node by equal outgoing spin currents to the right and
left \cite{Bauer:prep}. (As the interfacial scrambling is only partial
and the spacer is ballistic, the last statement should not be taken
literally, but as an effective theory which is valid after renormalizing the
interfacial conductance parameters.)
On typical FMR time scales, this process occurs practically instantaneously.
The net spin torque at one interface is therefore
just the difference of the pumped spin currents divided by two:
$\text{\boldmath$\tau$}_1=(\mathbf{I}_{s2}^{\text{pump}}-\mathbf{I}_{s1}^{\text{pump}})/2=-\text{\boldmath$\tau$}_2$.
When one ferromagnet is stationary, see the left drawing in
Fig.~\ref{fig2}, the dynamics of the other film, \textit{Fi}, is
governed by the LLG equation with a damping parameter
$\alpha_i=\alpha_i^{(0)}+\alpha_i^\prime$ enhanced with respect to
the intrinsic value $\alpha_i^{(0)}$ by
$\alpha_i^\prime=\gamma\hbar g^{\uparrow\downarrow}/(8\pi\mu_i)$,
where $\mu_i$ is the total magnetic moment of \textit{Fi}. Since
$\mu_i$ scales linearly with the volume of the ferromagnet and
$g^{\uparrow\downarrow}$ scales with its interface area,
$\alpha_i^\prime$ is inversely proportional to the film thickness.

When both magnetizations are allowed to precess, see the right drawing in Fig.~\ref{fig2}, the LLG equation expanded to include the spin torque reads
\begin{eqnarray}
\frac{d\mathbf{m}_i}{dt}&=&-\gamma\mathbf{m}_i\times\mathbf{H}^i_{\text{eff}}+\alpha_i^{(0)}\mathbf{m}_i\times\frac{d\mathbf{m}_i}{dt}\nonumber\\
&&+\alpha_i^\prime\left[\mathbf{m}_i\times\frac{d\mathbf{m}_i}{dt}-\mathbf{m}_j\times\frac{d\mathbf{m}_j}{dt}\right]\,,
\label{em}
\end{eqnarray}
where $j=1(2)$ if $i=2(1)$. As a simple example, consider a system in the parallel configuration, $\mathbf{m}_1^{(0)}=\mathbf{m}_2^{(0)}$, with matched resonance conditions. In addition, let us assume the resonance precession is circular. If we linearize Eq.~(\ref{em}) in terms of small deviations $\mathbf{u}_i(t)=\mathbf{m}_i(t)-\mathbf{m}_i^{(0)}$ of the magnetization direction $\mathbf{m}_i$ from its equilibrium value $\mathbf{m}_i^{(0)}$, we find that the average magnetization deviation $\mathbf{u}=(\mathbf{u}_1\mu_1+\mathbf{u}_2\mu_2)/(\mu_1+\mu_2)$ is damped with the intrinsic Gilbert parameter $\alpha^{(0)}$, whereas the difference $\Delta\mathbf{u}=\mathbf{u}_1-\mathbf{u}_2$ relaxes with enhanced damping constant $\alpha=\alpha^{(0)}+\alpha_1^\prime+\alpha_2^\prime$.

\begin{figure*}
\includegraphics[width=17.9cm,clip=]{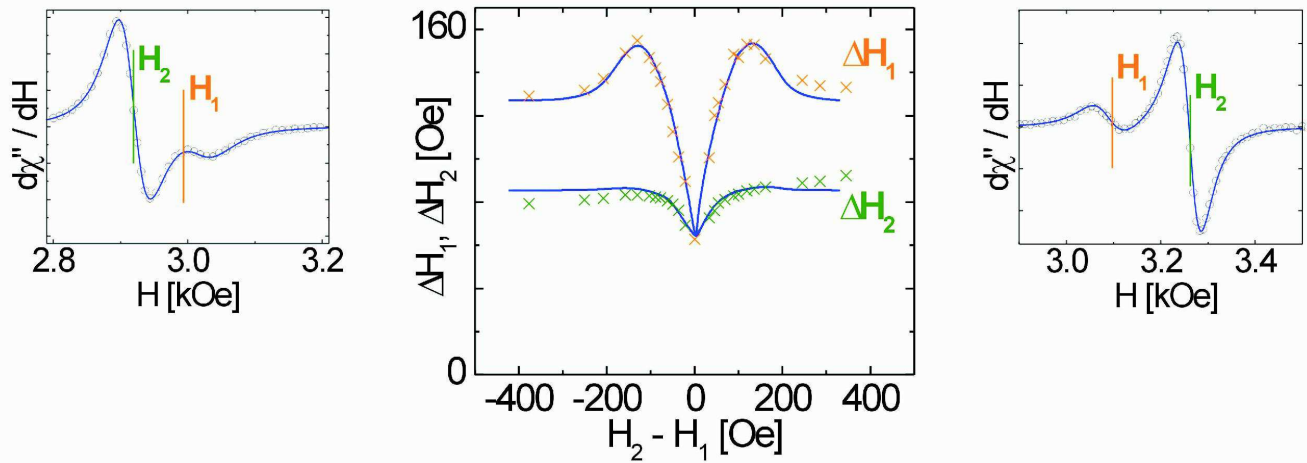}
\caption{\label{fig3} Comparison of theory (solid lines)  with
RT measurements (symbols) close to and at the crossover of the FMR
fields, marked by the shaded area in Fig.~\ref{fig1}. The left and
right frames show FMR signals for the field difference, $H_2-H_1$,
of -78~Oe and +161~Oe, respectively. The theoretical results are
parameterized by the full set of magnetic parameters which were
measured independently \cite{Urban:prl01}. The magnitude of the
spin-pump current was determined by the line width at large
separation of the FMR peaks. The middle frame displays the
effective FMR line width of magnetic layers for the signals fitted
by two Lorentzians as a function of the external field. At
$H_1=H_2$, the FMR line widths reached their minimum values at the
level of intrinsic Gilbert damping of isolated films. The
calculations in the middle frame did not take small variations of
the intrinsic damping with angle $\varphi$ into account, which
resulted in deviations between theory and experiment for larger
$|H_1-H_2|$. Note that $\Delta H_1$ first increases before
attaining its minimum, which is due to excitation of the
antisymmetric collective mode.}
\end{figure*}

Measuring the spin torques requires independent control of the
precessional motion of the two \textit{F} layers, with FMR
absorption line widths of isolated films dominated by the
intrinsic Gilbert damping. Both conditions were met by
high-quality crystalline Fe(001) films grown on 4x6 reconstructed
GaAs(001) substrates by Molecular Beam Epitaxy
\cite{Urban:prl01,Enders:jap01}. Fe(001) films were deposited at
room temperature (RT) from a thermal source at a base pressure of less
then $2\times10^{-10}$~Torr and the deposition rate was
$\sim1$~ML/min. For the experiments discussed below, single Fe
ultrathin films with thicknesses $d_F=$~11,16,21,31~ML were grown
directly on GaAs(001) and covered by a 20~ML protective Au(001)
cap layer. The magnetic anisotropies as measured by FMR are
described by a constant bulk term and an interface contribution
inversely proportional to $d_F$. The Fe ultrathin films grown on
GaAs(001) and covered by gold have magnetic properties nearly
identical to those in bulk Fe, modified only by sharply defined
interface anisotropies. The in-plane uniaxial anisotropy arises
from electron hybridization between the As dangling bonds and the
iron interface atoms. These Fe films were then regrown as one
element of a magnetic bilayer structure and in the
following referred to as \textit{F1} layers. They were separated
from a thick Fe layer, \textit{F2}, of 40~ML thickness by a 40~ML
Au spacer. The magnetic bilayers were covered by
20~ML of protective Au(001). The complete structures are therefore
GaAs/Fe(8,11,16,21,31)/40Au/40Fe/20Au(001), where the integers
represent the number of MLs.  The electron mean free path in thick
films of gold is 38~nm \cite{Enders:jap01} and, consequently, the
spin transport even in the 40~ML (8~nm) Au spacer is purely
ballistic. The interface magnetic anisotropies allowed us to
separate the FMR fields of the two Fe layers with
resonance-field differences that can exceed 5 times the FMR
line widths, see Fig.~\ref{fig1}. Hence, the FMR measurements for
\textit{F1} in double layers can be carried out with a nearly
static \textit{F2}.

The FMR line width of \textit{F1} increases in the presence of
\textit{F2}. The difference $\Delta H^\prime$ in the FMR line
widths between the magnetic bilayer and single-layer
structures is nearly inversely proportional to the thin-film
thickness
 $d_F$ \cite{Urban:prl01}, proving that $\Delta H^\prime$ originates at the \textit{F1{\rm /}N} interface. Secondly, $\Delta H^\prime$ is linearly dependent on microwave frequency for both the in-plane (the saturation magnetization parallel to the film surface) and perpendicular (the saturation magnetization perpendicular to the film surface) configurations, strongly implying that the additional contribution to the FMR line width can be described strictly as an interface Gilbert damping \cite{Urban:prl01}.
At the FMR, the film precessions are driven by an applied rf field. When the resonance fields are different, one layer (say \textit{F1}) is at
 resonance with maximum precessional amplitude while the other layer (\textit{F2}) is off resonance with small precessional amplitude, see Fig.~\ref{fig2}.
  The spin-pump current for \textit{F1} reaches its maximum while \textit{F2} does not emit a significant spin current at all. \textit{F2} acts as a spin sink causing the nonlocal damping for \textit{F1}. The \textit{N{\rm /}F2} interface provides a \textquotedblleft spin-momentum brake\textquotedblright\ for the \textit{F1} magnetization. The corresponding additional Gilbert parameter $\alpha^\prime$ for a 16~ML Fe is significant, being similar in magnitude to the intrinsic Gilbert damping in isolated Fe films, $\alpha^{(0)}=0.0044$.

These assertions can be tested by employing the in-plane
uniaxial anisotropy in \textit{F1} to intentionally tune the
resonance fields for \textit{F1} and \textit{F2} into a crossover
which is shown in the shaded area of Fig.~\ref{fig1}.
When the resonance fields are identical, $H_1=H_2$,
the rf magnetization components
of \textit{F1} and \textit{F2} are parallel to each other, see the
right drawing in Fig.~\ref{fig2}.
The total spin currents across the \textit{F1{\rm /}N}
and \textit{N{\rm /}F2} interfaces therefore vanish resulting in zero excess
damping for \textit{F1} and \textit{F2}, see Eq.~(\ref{em}), which is
experimentally verified, as shown in Fig.~\ref{fig3}.
For a theoretical analysis, we solved
Eq.~(\ref{em}) and determined the total FMR signal as a function
of the difference between the resonance fields $H_2-H_1$. The
theoretical predictions are compared with measurements in
Fig.~\ref{fig3}. The
remarkable good agreement between the experimental results and
theoretical predictions provides strong evidence that the dynamic
exchange coupling not only contributes to the damping but leads to
a new collective behavior of magnetic hybrid structures.

We have additionally carried out our measurements on
samples with Au spacer thickness between 14 and 100 monolayers.
The weak dependence of the FMR response on the spacer thickness
fully supports our picture of the long-ranged dynamic interaction.

In conclusion, we found decisive experimental and theoretical
evidence for a new type of exchange interaction between
ferromagnetic films coupled via normal metals. In contrast to the
well-known oscillatory exchange interaction in the ground state,
this coupling is dynamic in nature and long ranged. Precessing
magnetizations feel each other through the spacer by exchanging
nonequilibrium spin currents. When the resonance
frequencies of the ferromagnetic banks differ, their motion
remains asynchronous and net spin currents persist. However, when
the ferromagnets have identical resonance frequencies, the
coupling quickly synchronizes their motion and equalizes the spin
currents. Since these currents flow in opposite directions, the
net flow across both \textit{F1{\rm /}N} and \textit{N{\rm /}F2}
interfaces vanishes in this case. The lifetime of the arising
collective motion is limited only by the intrinsic local damping.
These effects can be well demonstrated in FMR measurements.

We are grateful to B.~I. Halperin for stimulating discussions. This
work was supported in part by the NEDO International Joint
Research Grant Program \textquotedblleft
Nano-magnetoelectronics\textquotedblright, NSF Grant DMR 02-33773,
and the FOM.

\end{document}